\documentclass[12pt]{article}
\usepackage{amsmath,amssymb,epsfig}
\usepackage{cancel}
\usepackage{color}
\usepackage{bm}
\usepackage{braket}
\usepackage{ulem}

\makeatletter

\@addtoreset{equation}{section}
\makeatother

\newcommand{\sgn}{\text{sgn}}

\begin{document}

\title{
\baselineskip 14pt
\hfill \hbox{\normalsize EPHOU-19-009}\\ 
\hfill \hbox{\normalsize WU-HEP-19-05} \\
\hfill \hbox{\normalsize KUNS-2766}
\vskip 1.7cm
\bf
Revisiting instabilities of $S^1/Z_2$ models with loop-induced Fayet-Iliopoulos terms
\vskip 0.5cm
}
\author{
\centerline{Hiroyuki~Abe$^{1}$, \
Tatsuo~Kobayashi$^{2}$, \
Shohei~Uemura$^{3}$ and \
Junji~Yamamoto$^{4}$}
\\*[20pt]
\\
{\it \normalsize 
\centerline{${}^{1}$Department of Physics, Waseda University, Tokyo 169-8555, Japan}}
\\
{\it \normalsize 
\centerline{${}^{2}$Department of Physics, Hokkaido University, 
Sapporo 060-0810, Japan}}
\\
{\it \normalsize 
\centerline{${}^{3}$CORE of STEM, Nara Women's University, Nara 630-8506, Japan}}
\\
{\it \normalsize 
\centerline{${}^{4}$Department of Physics, Kyoto University, 
Kyoto 606-8502, Japan}}
\\*[50pt]}

\date{
\centerline{\small \bf Abstract}
\begin{minipage}{0.9\linewidth}
\medskip 
\medskip 
\small
We study Fayet-Iliopoulos (FI) terms of 5-dimensional supersymmetric $U(1)$ gauge theory compactified on $S^1/Z_2$.
In this model, loop diagrams including matter hypermultiplets and brane chiral multiplets
induce FI-terms localized at the fixed points.
Localized FI-terms lead instabilities of bulk modes.
The form of the induced FI-terms strictly depends on wave function profiles of matter multiplets.
It is a non-trivial question whether the vacuum of 1-loop corrected potential
is stable under  radiative corrections.
We investigate this issue and it is found that the stable configuration is obtained when the bulk zero modes shield the brane charge completely.
\end{minipage}
}

\newpage

\begin{titlepage}
\maketitle
\thispagestyle{empty}
\clearpage
\end{titlepage}

\section{Introduction}

Supersymmetric gauge theories in higher dimensions have been studied for a long time.
It is an attractive extension of the standard model from phenomenological perspective including hierarchy, flavor, and grand unification \cite{Randall:1998uk, Abe:2008sx}.
It is also motivated by quantum theory of gravity
since superstring is defined on 10 dimensional spacetime and is supersymmetric.
To realize our standard model universe from such theories,
we must compactify the extra dimensions and need supersymmetry (SUSY) breaking.
The Fayet-Iliopoulos term (FI-term) of Abelian gauge theory
is a possible source of SUSY breaking \cite{Fayet:1974jb}.
It is a gauge invariant linear term of the auxiliary component of the $U(1)$ vector multiplet.
It is invariant under gauge and global SUSY transformation,
but it is not invariant under local SUSY unless the $U(1)$ is related to the gauged $U(1)_R$
\cite{Barbieri:1982ac, Binetruy:2004hh}\footnote{Another procedure without gauged $U(1)_R$ is proposed recently \cite{Cribiori:2017laj}.}
or collaborates with the Green-Schwarz like anomaly cancellation mechanism \cite{Green:1984sg}.
However, in higher dimensional theories, the situation is different.
Some sort of FI-terms can be consistent even in supergravity and they are localized at branes or fixed points of orbifolds \cite{Barbieri:2002ic, ArkaniHamed:2001tb}.
In that case, although tree level FI-term does not exist, radiative corrections induce the FI-terms \cite{Ghilencea:2001bw}.

FI-terms in higher dimensional theory can play another important role for phenomenology.
The localized FI-term generates a local potential of the real scalar 
component of bulk vector multiplet $\Sigma$, and $\Sigma$ develops its vacuum expectation value (VEV), 
which corresponds to  
a kink mass on the compact space.
It leads instabilities of bulk fields, especially localization of bulk zero modes at the fixed points in th 
$S^1/Z_2$ orbifold 
\cite{ArkaniHamed:2001tb, GrootNibbelink:2002qp, GrootNibbelink:2002wv, Abe:2002ps} and the $T^2/Z_2$ orbifold \cite{Lee:2003mc}.
Interactions in 4d low-energy effective theories are given by overlap integral of bulk wave functions.
Such localized modes imply a hierarchical structure of 4d couplings. 
It may be the origin of the observed flavor structure \cite{Kaplan:2001ga, Marti:2002ar, Abe:2018ylo}.

Instabilities of bulk fields have further effects for the vacuum structure of the orbifold models.
The computation of the radiative corrections to the FI-term depends on
the mode expansion of the bulk fields.
It is natural to expect that new mode expansion induces new quantum corrections.
It may change 1-loop effective potential.
Therefore, instability of the bulk fields may imply the instability of the 1-loop corrected vacuum of the scalar fields.
We should reconsider the localized FI-terms on the orbifold.
That is our purpose on this paper.

This paper is organized as follows.
In section 2, we review localized FI-terms induced by quantum corrections.
The localized FI-term induces nonzero VEV of $\Sigma$.
It affects equations of motion for bulk fields and their wave function profiles.
In Section 3, we consider the localized FI-term induced by quantum corrections again.
We find that stability of 1-loop corrected $\braket{\Sigma}$ background depends on configuration of the brane charges and zero mode profiles.
It is necessary for the stable configuration that the zero modes are localized at the fixed points and they shield the brane charges completely, otherwise induced FI-term shifts the $\braket{\Sigma}$ background again.
Section 4 is our conclusion.
In appendix, we study wave function profiles of parity odd modes and summation of massive mode wave 
functions explicitly.

\section{Localized FI-terms on orbifolds}

\begin{figure}[tbph]
\centering
\epsfig{file=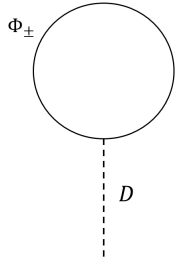,scale=0.15}
\caption{Tadpole diagram contributing to the FI-term.}
\label{fig:tadpole}
\end{figure}

In this section, we briefly review the localized FI-term of supersymmetric Abelian gauge theory
induced by quantum corrections.
We consider 5d $\mathcal{N}=1$ SUSY gauge theory compactified on $S^1/Z_2$.
$Z_2$ acts on $S^1$ as $y \rightarrow -y$, where $y$ is the coordinate of $S^1$, and $y\sim y+2\pi R$.
There are two fixed points at $y=0$ and $y=\pi R$.
This theory contains a gauge multiplet $V=(A_M, \lambda^i, \Sigma, \vec{D})$, and a hypermultiplet $H=(\phi_\pm, \psi, F_\pm)$.
The gauge multiplet contains 5d gauge field $A_M$, gauginos $\lambda^i$, one real scalar $\Sigma$ and a triplet auxiliary field $\vec D$.
The hypermultiplet contains two complex scalar fields $\phi_\pm$, one Dirac spinor $\psi$ and two auxiliary fields $F_\pm$.
$H$ is charged under the $U(1)$.
Its charge is denoted by $q$.
The
bulk Lagrangian is given by  \cite{ArkaniHamed:2001tb}:
\begin{align}
\mathcal{L}_{{\rm bulk}}= &-\frac 1 2 (\partial_M \Sigma)^2 +\frac 1 2 \mathcal{D}_M \Phi_+^\dagger \mathcal{D}^M \Phi_+ - \mathcal{D}_M \Phi_-^\dagger \mathcal{D}^M \Phi_- + i \bar \psi \gamma^M \mathcal{D}_M \psi \nonumber \\
&-gq D^3\left( \Phi_+^\dagger \Phi -\Phi_-^\dagger \Phi_- \right) -gq \left( \left(D^1 - iD^2\right)\Phi_+^T \Phi_- + h.c. \right) \nonumber \\
&-g^2 q^2 \Sigma^2 \left(\Phi_+^\dagger \Phi_+ +\Phi_-^\dagger \Phi_- \right)
-gq\Sigma \bar \psi \psi+...
\label{eq:bulk_action}
\end{align}
where $\mathcal{D}_M$ is the covariant derivative.
Parity assignment of the bulk fields preserving 4d $\mathcal{N}=1$ supersymmetry
and $U(1)$ gauge symmetry is given in \cite{Ghilencea:2001bw}, and it is shown in Table \ref{tab:parity}.
\begin{table}[htb]
\begin{center}
\begin{tabular}{|c|c|c|c|c|c|c|c|c|c|c|} \hline
	& $A_\mu$ & $A_5$ & $\lambda^1$ & $\lambda^2$ & $\Sigma$ & $D^{1,2}$ & $D^3$
	& $\phi_\pm$ & $\psi$ & $F_\pm$ 
\\ \hline 
	parity & + & $-$ & $+$ & $-$ & $-$ & $-$ & +& $\pm$ & $\gamma_5$ & $\pm$
\\
\hline
\end{tabular}
\caption{Parity assignment of the bulk fields.}
\label{tab:parity}
\end{center}
\end{table}
This notation implies that the 4d D-term is denoted by $D=D^3 - \partial_y \Sigma$.
Although the tree level Lagrangian does not have the FI-term,
quantum corrections can induce it.
The quantum corrections to the coefficient of $D^3$
is represented by tadpole diagrams shown in 
Fig.\ref{fig:tadpole} \cite{Ghilencea:2001bw, Barbieri:2000vh}.
In the present paper, we concentrate only on the coefficient of $D^3$.
The coefficient of $\partial_y \Sigma$ is also given by a similar diagram,
but the fields running in the loop are replaced by $\psi$.
They are the same contribution and it is sufficient for our purpose to consider either of them if SUSY is not broken.
The effective FI-term is calculated as
\begin{align}
\int \frac{d^4 p_4}{(2\pi)^4} \frac{gq}{p_4^2-m^2} \sim \frac{gq}{16\pi^2}
(\Lambda^2 - \frac{1}{4}m^2 \ln \Lambda^2),
\end{align}
up to finite terms,
where $m$ is the 4d mass of scalar fields in the loop and $\Lambda$ is the UV cutoff.

Compactifying the 5th direction,
mode expansions of $\phi_\pm$ are given by \cite{GrootNibbelink:2002qp},
\begin{align}
\phi_+(x, y) &= \sum_{n \geq 0} \eta_n \phi_+^n(x) \cos \left( \frac{n} R y \right),
\label{eq:phi_+}
\\
\phi_-(x, y) &= \sum_{n \geq 1} \eta_n \phi_-^n(x) \sin \left( \frac{n} R y \right),
\label{eq:phi_-}
\end{align}
where $\eta_n$ are the normalization factors, i.e.,
$\eta_0 = 1/\sqrt{\pi R}$ and  $\eta_{n \geq 1} = \sqrt{2/\pi R}$.
We normalize the fields on $[0,\pi R]$.
The 5-dimensional Lagrangian is reduced to,
\begin{align}
\mathcal{L}_{{\rm bulk}} \supset gq D^3 
\sum_{n \in N}
\eta_n^2\left(
|\phi_+^n(x)|^2 \cos^2 \left(\frac {n y} R \right) - |\phi_-^n(x)|^2 \sin^2 \left(\frac {n y} R\right)
\right).
\end{align}
Thus the quantum corrections to the FI-term generated by bulk fields are given by 
\begin{align}
\xi_{{\rm bulk}}(y) =& \sum_{n} \frac{gq}{16\pi^2} (\Lambda^2 - \frac 1 4 \frac{n^2}{R^2}  \ln \Lambda^2 ) \left( \cos^2 \left(\frac {n y} R \right) -  \sin^2 \left(\frac {n y} R\right)
\right).
\label{eq:fcn_ser}
\end{align}
This absolutely diverges at the fixed points\footnote{
The function series of (\ref{eq:fcn_ser}) does not converge absolutely, but it conditionally converge except for the fixed points.
It is always possible to shift the convergence value of conditionally convergent series arbitrarily by changing its order.
In this paper, we assume the FI-term is not derived except for the fixed points
since the bulk FI-term can violate either of the 4d SUSY or 5d Lorentz invariance.} and it can not be removed. 
Therefore, the induced FI-term becomes Dirac delta function,
\begin{align}
\xi_{{\rm bulk}}(y) =& \frac{gq}{16\pi^2} (\Lambda^2 + \frac 1 4 \ln \Lambda^2 \partial_y^2) \left(\delta(y)+ \delta(y-\pi R)\right).
\label{eq:xi_bulk}
\end{align}
Hence, the localized FI-term is obtained \cite{Ghilencea:2001bw, GrootNibbelink:2002qp, GrootNibbelink:2002wv}
.

In addition, brane multiplets confined at the fixed points can contribute to $\xi$, too.
We introduce two chiralmultiplets $C_0$ and $C_\pi$ localized at the fixed points.
$C_0$ is localized at $y=0$ and $C_\pi$ is localized at $y=\pi R$.
They are coupled to the gauge multiplet and their charges are $q_0$ and $q_\pi$ respectively.
Their couplings to the gauge multiplets include
\begin{align}
\mathcal{L}_{{\rm brane}} \supset \sum_{I=0,\pi} g q_I (D^3-\partial_y \Sigma) \phi_I^\dagger \phi_I,
\end{align}
where $\phi_0$ and $\phi_\pi$ are the scalar components of $C_0$ and $C_\pi$.
They contribute to the localized FI-terms as,
\begin{align}
	\xi_{{\rm brane}}(y) =g \frac{\Lambda^2}{16\pi^2}\sum_{I=0,\pi}\delta(y-IR) q_I.
\end{align}
The quantum corrections to the localized FI-term are summarized: 
\begin{align}
	\xi(y) = \sum_{I=0,\pi} (\xi_I + \xi_I^{\prime \prime} \partial^2_y)\delta(y-IR),~~
	\xi_I = g \frac{\Lambda^2}{16\pi^2}(\frac{q}{2}+q_I),\, \,\,\,\,\,
	\xi_I^{\prime \prime} = \frac{gq }{128 \pi^2} \ln \Lambda^2.
\label{FI-term}
\end{align}
Here, $\xi_0 + \xi_\pi$ is always zero and the 4d FI-term $\int \xi(y) dy$ is  zero if the sum of the $U(1)$ charges $q+q_0 +q_\pi$ is set to zero.
Moreover, $\xi_0''$ is equal to $\xi''_\pi$.

In this section, we have considered models containing a single bulk hypermultiplet and a single chiral multiplet at each fixed point.
It is straightforward to generalize the result to models of several bulk hypermultiplets and brane superfields.
In such a case, the charges are replaced by trace of charge operators,
and it is written as
\begin{align}
\xi_I = g \frac{\Lambda^2}{16\pi^2}(\frac{1}{2}\text{tr}(q)+\text{tr}(q_I)),\, \,\,\,\,\,\xi_I^{\prime \prime} =\frac{g~\text{tr}(q)}{128 \pi^2}\ln \Lambda^2,	
\label{eq:multiFI}
\end{align}
where the trace runs over all hypermultiplets and brane chiral multiplets. 


\subsection{Localization of Bulk Fields}

Now, we study effects of the localized FI-terms on wave function profiles of the bulk scalar fields.
Since the bulk scalar fields $\Sigma$, $\phi_-$ are parity odd and $\phi_+$ is even under the orbifolding,
boundary conditions of these fields are given by
\begin{align}
\Sigma(y=0) =\Sigma(y=\pi R) &= \phi_-(y=0) = \phi_-(y=\pi R) =0,
\nonumber
\\
\partial_y \phi_+ (y=0)&=  \partial_y \phi_+ (y=\pi R)= 0.
\nonumber
\end{align}

The field $\Sigma$ develops its VEV along the $D$-flat direction \cite{GrootNibbelink:2002qp},
\begin{align}
\partial_y \braket{\Sigma} = \xi(y)+C,
\label{eq:D-flat}
\end{align}
where $C$ is a constant term proportional to the sum of $U(1)$ charges.
If the sum is set to zero, $C$ vanishes.
Then, equations of motion of the bulk scalar
$\phi_{\pm}$ are written as \cite{GrootNibbelink:2002qp}.
\begin{align}
\Delta_\pm \phi_\pm +\lambda \phi_\pm = 0,~~
\Delta_\pm = \partial_y^2 \mp gq \partial_y \braket{\Sigma} - g^2q^2 \braket{\Sigma}^2.
\label{eq:wave_equation}
\end{align}
If $\xi(y)$ and $C$ are equal to zero, the equation of motion is the normal harmonic oscillator,
and its solution is (\ref{eq:phi_+}) and (\ref{eq:phi_-}).
Otherwise wave function profiles are changed.

To show the effects of the FI-term,
we shortly review one example studied in \cite{GrootNibbelink:2002qp}.
Suppose
$\xi(y) = \xi_0'' \partial_y^2 \delta(y) +\xi_0'' \partial_y^2 \delta(y-\pi R)$, where $\xi''_0$ is a  constant and $q \xi_0''>0.$
The solution of D-term condition is given by
\begin{align}
\braket{\Sigma}(y) = \xi_0^{\prime \prime}
\Big( \partial_y \delta(y) +\partial_y \delta(y-\pi R) \Big).
\end{align}
Here, we concentrate on the zero mode.
The parity odd mode $\phi_-$ has no zero mode.
The zero mode of $\phi_+$ has delta function like profile, i.e.,
\begin{align}
|\phi_{+,0}|^2(y) = \frac 1 2 \left(\delta(y) +\delta(y-\pi R)\right).
\end{align}
The zero mode is localized at the fixed points.
It is drastically changed from the constant solution of (\ref{eq:phi_+}).

In the computation in \cite{GrootNibbelink:2002qp},
$\braket{\Sigma}$ is set to zero and the bulk scalars are decomposed by the simple harmonic oscillators; (\ref{eq:phi_+}) and (\ref{eq:phi_-}).
As the results, divergent contribution of quantum corrections to $\xi$ at the fixed points is obtained.
However, if the FI-term has nonzero value, 
the $\Sigma$ field  develops its VEV $\braket{\Sigma}$ and 
the bulk fields are not decomposed by simple functions such as (\ref{eq:phi_+}) and (\ref{eq:phi_-}).
Then, we naively expect that this localized FI-term changes the 1-loop corrections itself.
Thus, questions occur.
Are the 1-loop FI-term (\ref{FI-term}) and the $\Sigma$ background stable under the redefinition of the KK-reduction of the bulk field?
If not, is there any stable configuration where the bulk fields profile and $\xi$ are consistent?

We shall answer these questions in the next section.

\section{Corrected FI-terms}

In this section, we recompute quantum corrections to FI-terms by the bulk scalar fields
and investigate whether 1-loop corrected $\braket{\Sigma}$ background is stable under quantum corrections.
We consider models whose sum of the $U(1)$ charges is zero and the constant term $C$ in (\ref{eq:D-flat}) is zero.
For the moment, we also assume that only the $\Sigma$ field develops its VEV, but VEVs of other fields vanish.
If the charge sum does not vanish, the $U(1)$ gauge symmetry or 4d SUSY is broken down.
We briefly discuss such a case at the end of this section.

When the KK-decomposition of the bulk scalar fields are represented as
\begin{align}
\phi_+(x,y) = \sum_{\lambda} f_+^\lambda (x) \phi_+^\lambda (y),\\
\phi_-(x,y) = \sum_{\lambda} f_-^\lambda (x) \phi_-^\lambda (y),
\end{align}
where $\lambda$ is the eigenvalues of (\ref{eq:wave_equation}),
quantum corrections to the coefficients of $D^3$
are given by
\begin{align}
\xi(y) =& \sum_{\lambda} \frac{gq}{16\pi^2} (\Lambda^2 - \frac 1 4 \lambda \ln \Lambda^2)
(|\phi_+^\lambda|^2(y) - |\phi_-^\lambda|^2(y)) + \sum_{I=0,\pi}  \frac{g q_I}{16\pi^2} \Lambda^2 \delta(y-I R). 
\label{fom:xi}
\end{align}
To compute the FI-term, we must know wave functions and eigenvalues of the whole KK-modes.

In order to illustrate our story, we study simple symmetric case for the first step,
and next, we study more general asymmetric case.

\subsection{Symmetric Case}

Here, we consider the symmetric model, i.e. $q_0=q_\pi$.
We assume the sum of $U(1)$ charges is set to zero, and thus,
the bulk charge is twice as big as that of the localized charge: $q =-2 q_0$.
We assume the tree level Lagrangian has no FI-term
and $\braket{\Sigma} =0$ at the tree level.
The bulk wave function is given by (\ref{eq:phi_+}) and (\ref{eq:phi_-}),
and the induced localized FI-term is written as,
\begin{align}
	\xi_{{\rm 1-loop}} (y) &= (\xi_0 + \xi^{\prime \prime}_0 \partial^2_y )\delta(y) + (\xi_\pi+ \xi^{\prime \prime}_\pi \partial^2_y) \delta(y-\pi R),
\nonumber
\\
	\xi_0 &= \xi_\pi = 0,
\nonumber
\\
	\xi^{\prime\prime}_0&= \xi_\pi^{\prime\prime} =  \frac{gq}{64\pi^2 }\ln \Lambda^2.
\label{eq:xi_sym1}
\end{align}
Solving the D-flat condition $\partial_y \braket{\Sigma} = \xi(y),$
the 1-loop corrected $\braket{\Sigma}$ VEV is obtained
\begin{align}
	\braket{\Sigma}(y) = \xi_0^{\prime \prime} \Big(\partial_y \delta(y) +\partial_y \delta(y-\pi R)\Big).
\end{align}
We obtain the new VEV of $\Sigma$.
Then, we recompute the mode expansion of $\phi_\pm$.

The VEV $\braket{\Sigma}$ includes derivatives of the delta functions.
To solve the equation of motion (\ref{eq:wave_equation}),
we must regularize the Dirac delta function.
We adopt the regularization used in \cite{GrootNibbelink:2002qp},
\begin{align}
\delta_\rho(y) = 
\begin{cases}
\frac 1 {\rho^2} (y+\rho) &(-\rho < y < 0)\\
-\frac 1 {\rho^2} (y-\rho) &(0 \leq y < \rho)\\
0 &(\rho \leq |y|) ,
\end{cases}
\label{eq:reg}
\end{align}
where
$\delta_\rho(y) \rightarrow \delta(y)$ as $\rho \rightarrow 0$.
To simplify the Schr\"odinger equation (\ref{eq:wave_equation}), we rewrite the wave function as
\begin{align}
\phi_\pm (y) = \psi_\pm (y) \exp \left[ \pm gq \int_0^y \braket{\Sigma}(y^\prime) dy^\prime  \right].
\label{eq:eta}
\end{align}
Then, the wave equation for $\phi$ is rewritten by $\psi$.
\begin{align}
\psi''_\pm \pm 2gq \braket{\Sigma} \psi'_\pm +\lambda \psi_\pm=0.
\label{d_eom}
\end{align}
The parity even solution is given by
\begin{align}
\psi_{+} (y) =
\begin{cases}
	A_+ e^{(\omega+\sqrt{\omega^2-\lambda})y}+B_+ e^{(\omega-\sqrt{\omega^2-\lambda})y} ~~~&(0 \leq y \leq \rho)\\
	C_+ e^{i\sqrt{\lambda}y}+D_+e^{-i\sqrt{\lambda}y} &(\rho < y<\pi R -\rho )\\
	E_+ e^{(-\omega+\sqrt{\omega^2-\lambda})(y-\pi R)}+F_+ e^{(-\omega-\sqrt{\omega^2-\lambda})(y-\pi R)}&(\pi R -\rho \leq y \leq \pi R ),
\end{cases}
\end{align}
where $\omega = gq \xi_0''/\rho^2$.
Here,
$\omega$ is positive since it is proportional to $q^2$.
The coefficients are determined by the boundary and continuous conditions.
They are written as
\begin{align}
&\alpha_+ A_+ + \alpha_- B_+=0,
\nonumber
\\
&A_+ e^{\alpha_+ \rho} + B_+ e^{\alpha_- \rho} 
= C_+ e^{i\sqrt{\lambda} \rho} + D_+ e^{-i\sqrt{\lambda} \rho},
\nonumber
\\
&\alpha_+ A_+ e^{\alpha_+ \rho} + \alpha_- B_+ e^{\alpha_- \rho} 
= i\sqrt{\lambda} C_+ e^{i\sqrt{\lambda} \rho} -i\sqrt{\lambda}  D_+ e^{-i\sqrt{\lambda} \rho},
\nonumber
\\
& C_+ e^{i\sqrt{\lambda} (\pi R-\rho)} + D_+ e^{-i\sqrt{\lambda} (\pi R-\rho)}
= E_+ e^{\alpha_- \rho} + F_+ e^{ \alpha_+ \rho},
\nonumber
\\
& i\sqrt{\lambda} C_+ e^{i\sqrt{\lambda} (\pi R-\rho)} -i\sqrt{\lambda}  D_+ e^{-i\sqrt{\lambda} (\pi R-\rho)}
= - \alpha_- E_+ e^{ \alpha_- \rho} - \alpha_+ F_+ e^{ \alpha_+ \rho},
\nonumber
\\
&\alpha_- E_+ + \alpha_+ F_+ = 0,
\end{align}
where $\alpha_\pm = \omega \pm \sqrt{\omega^2 -\lambda}$, and the eigenvalue $\lambda$ is given by
\begin{align}
\frac{(X+i\sqrt{\lambda})^2}{(X-i\sqrt{\lambda})^2} e^{2i\sqrt{\lambda} (\pi R-2 \rho)}=1
\label{eq:eigen_v_sym}
\end{align}
where
\begin{align}
X=\alpha_+ \frac{e^{\alpha_+ \rho} - e^{\alpha_- \rho}}{e^{\alpha_+ \rho} - \frac {\alpha_+}{\alpha_-} e^{\alpha_- \rho}}.
\end{align}
Hence,
$|X|$ diverges as $\rho \rightarrow 0$, and the right equation of (\ref{eq:eigen_v_sym}) approaches to
\begin{align}
e^{i\sqrt{\lambda}\pi R} = \pm 1.
\end{align}
The eigenvalue $\lambda$ is approaching to $\frac {n^2 }{R^2}$ \cite{GrootNibbelink:2002qp}.
The boundary conditions are satisfied with
\begin{align}
A_+ &= \frac{-2i\sqrt{\lambda}}
{(e^{\alpha_+ \rho} - \frac {\alpha_+}{\alpha_-} e^{\alpha_- \rho})(X-i\sqrt{\lambda})}
e^{-i\sqrt{\lambda} \rho} D_+ ,
\nonumber
\\
B_+ &= -\frac {\alpha_+}{\alpha_-} A_+,
\nonumber
\\
C_+ &= - \frac {X + i \sqrt{\lambda}}{X-i\sqrt{\lambda}} e^{-2i \sqrt{\lambda F}\rho} D_+ ,
\nonumber
\\
E_+ &= -\frac{\alpha_+}{\alpha_-} F_+,
\nonumber
\\
F_+ &= \frac{2i\sqrt{\lambda}}
{(e^{\alpha_+ \rho} - \frac {\alpha_+}{\alpha_-} e^{\alpha_- \rho})(X + i\sqrt{\lambda})}
e^{-i\sqrt{\lambda} (\pi R-\rho)} D_+.
\label{eq:F_+}
\end{align}
This solution has several interesting futures.
For nonzero value of $\lambda$, the coefficients $A_+,B_+,E_+,F_+$ are approaching to $0$ in the limit
$\rho \rightarrow 0$.
Therefore, all massive modes can not penetrate to the fixed points,
and they are purely bulk modes.
On the other hands, when $\lambda = 0$, the situation is completely opposite.
The solution is given by a constant solution: $\psi_+^0 = {\rm const}$ for $\lambda =0$.
Note that the
full wave function is given by  (\ref{eq:eta}).
Since the exponential part in (\ref{eq:eta}) is suppressed seriously in the bulk of the orbifold,
the zero mode profile has nonzero value only around the fixed points and it becomes a localized mode.\footnote{The massive modes are also suppressed on the bulk space.
However, the suppression factor of the first and the fifth equation of (\ref{eq:F_+}) are stronger than that of (\ref{eq:eta}). Thus, the massive modes are excluded from the fixed points.}

The mode expansion of $\phi_-$ is similar to that of $\phi_+$.
For massive modes, eigenvalues of $\phi_-$ are the same as $\phi_+$, 
$\lambda = \frac {n^2}{R^2}.$
The wave function is simple cosine function;
$\phi_-^n  = \sqrt{2/\pi R} \cos (n y/R) $ apart from the fixed points.
At the fixed points, since it is odd parity, we obtain $\phi_-^n =0$.
$\phi_-$ has no normalizable solution for $\lambda=0$.
It is because constant solution is not allowed for $\psi_-$\footnote{For precise computation of $\phi_-$, see Appendix \ref{app:odd}.}.

Mode expansions of $\phi_\pm$ are summarized as follows:
\begin{align}
&\phi_+^{n>0} = \sqrt{\frac 2 {\pi R}} \sin \left( \frac n R y \right),
\nonumber
\\
&\phi_-^{n>0} =
\begin{cases}
\sqrt{\frac 2 {\pi R}} \cos \left( \frac n R y \right) &(0<y<\pi R),\\
0 &(y=0,\pi R),
\end{cases}
\nonumber
\\
&\phi_+^0 = \sqrt{\frac{\delta(y) + \delta(y-\pi R)}2}.
\end{align}
This solution may look a bit strange.
The square root of the delta function just means that the wave function is localized at the fixed points and canonically normalized.
$\phi_+^0$ always appear as its square of absolute value in the whole discussion of this paper.

Substituting the above solution $\phi_\pm^\lambda$ in (\ref{fom:xi}),
we obtain the 1-loop induced FI-term again.
Since massive modes of $\phi_+$ and $\phi_-$ become zero at the boundaries,
they can not contribute to the localized FI-term.
The massive modes on the bulk cancel each others again.
Thus, the  contribution is due to the only zero mode of $\phi_+$,
\begin{align}
\xi_{{\rm bulk}} = \frac{gq}{32\pi^2} \Lambda^2 \sum_{I=0, R}  \delta(y-\pi I).
\end{align}
The contribution of the brane fields is unchanged.
It is written as,
\begin{align}
\xi_{{\rm brane}} = \frac{g}{16\pi^2} \Lambda^2 \sum_{I=0, R} q_I \delta(y-\pi I).
\end{align}
Since we assumed the sum of $U(1)$ charges vanishes, it cancels the bulk contribution.
As the result, we obtain the quantum correction to the FI-term,
\begin{align}
\xi(y)=\xi_{{\rm brane}}+\xi_{{\rm bulk}} = 0.
\label{eq:xi_sym2}
\end{align}
The quantum correction vanishes.
It is because the brane charges are completely shielded by the bulk zero mode.
The D-term potential vanishes.

\subsection{Asymmetric Case}

Next, we consider an asymmetric case.
We assume $q_0 \neq q_\pi$ and $q+ q_0 +q_\pi=0$.
The 1-loop induced FI-term is calculated as
\begin{align}
	\xi_{{\rm 1-loop}}(y) = &(\xi_0 + \xi^{\prime \prime}_0 \partial^2_y )\delta(y) + (\xi_\pi+ \xi^{\prime \prime}_\pi \partial^2_y) \delta(y-\pi R),\nonumber \\
	\xi_0 =& -\xi_\pi=\frac{g(q_0-q_\pi)}{32 \pi^2 } \Lambda^2,
\nonumber
\\
	\xi''_0=& \xi_\pi^{\prime\prime} = \frac{gq}{128 \pi^2} \ln \Lambda^2 .
\label{eq:xi_asym1}
\end{align}
Solving the D-flat condition, $\braket{\Sigma}$ is obtained as follows,
\begin{align}
\braket{\Sigma} = \frac 1 2 \xi_0 \sgn(y) + \xi_0''(\partial_y \delta(y) +\partial_y \delta(y-\pi R)),
\label{eq:Sigma_asy}
\end{align}
where $\sgn(y)= +1$ for $0 \leq y \leq \pi R$, and $\sgn(y)=-1$ for $-\pi R< y <0$.
The derivatives of the delta functions are regularized as same as the previous subsection.
Mode expansion of the bulk scalar is parallel to that of the symmetric case.
For even field, the solution is given as
\begin{align}
\psi_+^\lambda(y) =
\begin{cases}
A_+ e^{[ (\omega -\sigma)+\sqrt{(\omega-\sigma)^2 -\lambda}] y} + B_
+ e^{[(\omega -\sigma)-\sqrt{(\omega-\sigma)^2 -\lambda} ] y}~~&(0<y<\rho),\\
C_+ e^{(-\sigma+\sqrt{\sigma^2 -\lambda}) y} + D_+ e^{(-\sigma -\sqrt{\sigma^2 -\lambda} )y} &(\rho<y<\pi R-\rho),\\
E_+ e^{[-(\omega + \sigma)+\sqrt{(\sigma+\omega)^2 \lambda} ] (y-\pi R)} +
F_+ e^{[-(\omega + \sigma)-\sqrt{(\sigma+\omega)^2 -\lambda}](y-\pi R)} 
&(\pi R -\rho <y< \pi R),
\end{cases}
\end{align}
where $\omega = gq \xi''_0/\rho^2$, $\sigma = gq \xi_0/2$.
Note that it is $\psi_+^\lambda$, and is not $\phi_+^\lambda$.
Similar to the previous section, its boundary condition is written as
\begin{align}
&\alpha_+ A_+ +\alpha_- B_+ = 0,
\nonumber
\\
&\gamma_+ E_+ +\gamma_- F_+ =0,
\nonumber
\\
&A_+ e^{\alpha_+ \rho} +B_+ e^{\alpha_- \rho} 
= C_+ e^{\beta_+ \rho} + D_+ e^{\beta_- \rho},
\nonumber
\\
&\alpha_+ A_+ e_+^{\alpha_+ \rho} + \alpha_- B_+ e^{\alpha_- \rho}
= \beta_+ C_+ e^{\beta_+ \rho} +\beta_- D_+ e^{\beta_- \rho},
\nonumber
\\
&C_+ e^{\beta_+ (\pi R -\rho)} +D_+ e^{\beta_- (\pi R -\rho)} 
= E_+ e^{-\gamma_+ \rho} + F_+ e^{-\gamma_- \rho},
\nonumber
\\
&\beta_+ C_+ e^{\beta_+ (\pi R -\rho)} +\beta_- D_+ e^{\beta_- (\pi R -\rho)}
= \gamma_+ E_+ e^{-\gamma_+ \rho} +\gamma_- F_+ e^{-\gamma_- \rho},
\end{align}
where $\alpha_\pm = (\omega - \sigma) \pm \sqrt{(\omega - \sigma)^2 -\lambda},$
$\beta_\pm = -\sigma \pm \sqrt{\sigma^2 -\lambda},$
and $\gamma_\pm = -(\omega + \sigma)\pm \sqrt{(\omega + \sigma)^2 -\lambda}.$
It is trivial that $B_+/A_+ = -\alpha_+/\alpha_-$ and $F_+/E_+ =  -\gamma_+/\gamma_-$,
For $\lambda > 0$, its solution is obtained as follows,
\begin{align}
&A_+ =\frac {2\sqrt{\sigma^2 -\lambda} e^{\beta_- \rho}}
{
\alpha_+ (e^{\alpha_+\rho} -e^{\alpha_- \rho} )-\beta_+ ( e^{\alpha_+ \rho}- \frac{\alpha_+}{\alpha_-} e^{\alpha_- \rho} )
}
D_+,
\nonumber
\\
&C_+=
-
\frac {
\beta_- 
(
\gamma_- e^{-\gamma_+ \rho} - \gamma_+ e ^{-\gamma_- \rho}
)
-\gamma_+ \gamma_- 
(e^{-\gamma_+ \rho} - e ^{-\gamma_- \rho})
}
{
\beta_+ (
\gamma_- e^{-\gamma_+ \rho} - \gamma_+ e ^{-\gamma_- \rho}
)
-\gamma_+ \gamma_- 
(e^{-\gamma_+ \rho} - e ^{-\gamma_- \rho})
}
e^{-2\sqrt{\sigma^2-\lambda}(\pi R-\rho)}
D_+,
\nonumber
\\
&E_+ = - \frac {2\sqrt{\sigma^2 -\lambda} e^{\beta_- (\pi R -\rho)}}
{
\gamma_+ (e^{-\gamma_+\rho} -e^{-\gamma_- \rho} )-\beta_+ ( e^{-\gamma_+ \rho}- \frac{\gamma_+}{\gamma_-} e^{-\gamma_- \rho} )
}
D_+.
\end{align}
In the limit $\rho \rightarrow 0$, $\alpha_+ \rho$ and $-\gamma_- \rho$ diverge to $\infty$.
Its eigenvalue equation is obtained by $e^{-2\sqrt{\sigma^2 -\lambda} \pi R} =1$.
Then, we find that 
$\lambda\rightarrow \sigma^2+n^2/R^2$ as the limit  $\rho \rightarrow 0$ \cite{GrootNibbelink:2002qp}.
It is also similar to the symmetric case that $A_+, B_+, E_+,F_+ \rightarrow 0$, but $C_+ + D_+ \rightarrow 0$ as $\rho \rightarrow 0$.
The massive modes go to zero at the fixed points, and it is purely bulk mode.
For $\lambda = 0$, we have flat solution: $\psi_+^0 = {\rm const}$.
The full wave function is given by (\ref{eq:eta}).
Its exponential part is exponentially suppressed except for the fixed points.
Thus, the zero mode is localized at the fixed points, too.
It is almost parallel to the zero mode profile of the previous symmetric case.
However, there is one difference.
Because of the asymmetric charge configuration, $\braket{\Sigma}$ is not symmetric: 
$\braket{\Sigma}(0)\neq \braket{\Sigma}(\pi R)$.
The ratio of  the zero mode wave function values between two fixed points is
\begin{align}
\phi_+^0(0) : \phi_+^0(\pi R) = 1 : e^{\sigma \pi R}.
\end{align}
Thus, as $\rho \rightarrow 0$, we obtain the following asymmetric zero mode profile:
\begin{align}
\phi_+^0 = 
\sqrt{
\frac{1}{1+e^{2\sigma \pi R}} \delta(y) + \frac{e^{2\sigma \pi R}}{1+e^{2\sigma \pi R}} \delta(y-\pi R)
}.
\end{align}
The square root just means that $\phi_+^0$ is localized at the fixed points
and it is canonically normalized.

The mode expansion of parity odd modes $\phi_-$ is parallel to that of the symmetric case.
We can check that there is no zero mode for $\phi_-$.
The eigenvalues are the same as those of the even modes; $\lambda = \sigma^2 +n^2/R^2$.
The massive mode of $\phi_-$ is written as
\begin{align}
	\phi_-^{n>0}(y) =
	\begin{cases}
	\sqrt{\frac{2}{\pi R}} \sin(\frac{n}{R}y+\theta_n) &(0< y < \pi R ),\\
	0 &(y=0,\pi R),
	\end{cases}
\end{align}
where the constant phase $\theta_n$ is given by
\begin{align}
e^{i\theta_n} = \frac{-\sigma +i n/R}{|-\sigma +in/R|}.
\end{align}
Since all the massive modes are zero at the fixed points,
they can not contribute to the localized FI-term.
The cancellation of the massive modes except for the fixed points is more subtle than that of the symmetric case.
We do not consider it more deeply here, but we just assume that the bulk FI-term vanishes.
Substituting the above solution to (\ref{fom:xi}),
we obtain the one loop FI-term:
\begin{align}
&\xi(y) = \frac{g}{16 \pi^2} \Lambda^2 \left(\xi_1 \delta(y) -\xi_1 \delta(y-\pi R)
\right),
\nonumber
\\
&{\rm where}~~\xi_1
=\frac{-q_0 - q_\pi}{1+e^{2 \sigma \pi R}}+ q_0.
\label{eq:xi_asym2}
\end{align}
Since massive modes can not contribute to the localized FI-term,
the derivative term vanishes.
The FI-term has not vanished yet.
The vacuum configuration $\braket{\Sigma}$ would receive 1-loop correction again.
Our vacuum (\ref{eq:Sigma_asy}) is not stable under quantum corrections.

\subsection{Stable Configuration}

Our analysis contains two steps.
At the first step, we compute quantum corrections to the FI-term based on the mode expansion
derived by the tree level Lagrangian.
We assume 
$\braket{\Sigma}$ background is zero at the tree level.
The wave function profiles of the bulk scalar fields imply that all massive and zero modes spread on the bulk of the orbifold.
The quantum correction to the FI-term is contributed from all the modes,
but they cancel each other almost everywhere on the compact space.
Remaining corrections only appear at the fixed points.
We obtain the localized FI-terms.
At the second step, we reconsider quantum corrections to the FI-term.
In this step, we use the VEV of $\Sigma$ determined by the 1-loop FI-term.
Such a FI-term shifts $\braket\Sigma$ and mode expansions of the bulk fields are changed.
Zero mode is localized at the fixed points
and massive modes are confined in the bulk of the compact space except the fixed points.
For the symmetric case,
the 1-loop correction to FI-term has completely disappeared.
It is because the brane charges are shielded by the bulk zero mode and the massive modes cancel each others.
Since there is no source of the D-term potential,
VEV would not receive more corrections.
The 1-loop corrected background is the stable vacuum.
However, it is not true for the asymmetric case.
The 1-loop correction is not canceled completely.
$\braket \Sigma$ background is not stable yet.
One may expect that we can reach a stable configuration by repeating the above steps until FI-term vanishes.
However, the situation is more subtle.
Zero mode localization is caused by the derivatives of the delta functions in the FI-term.
Since the 1-loop FI-term (\ref{eq:xi_asym2}) does not include
derivatives, zero mode is no longer localized.
Thus, the 1-loop FI-term is contributed from all the massive modes.
Such a mode expansion leads derivative terms in the FI-term again\footnote{For precise computation, see Appendix \ref{app:xi''=0}.}.
It is not clear whether we have a consistent solution or not for the asymmetric case.

If the Lagrangian has a special value of FI-terms at the tree level,
we can find a stable configuration.
For instance, for the symmetric model with $q_0=q_\pi$, suppose the tree level FI term is written as
\begin{align}
\xi_{{\rm tree}} (y) &= \xi^{\prime \prime} \partial^2_y \delta(y) +\xi^{\prime \prime} \partial^2_y \delta(y-\pi R),
\end{align}
where $\xi''$ is an arbitrary constant such that $q \xi''>0$.
Then $\Sigma$ develops its VEV to cancel $\xi_{\rm tree}$.
In this case, the1-loop correction is the same as that of the symmetric case and it is zero.
Thus, 1-loop correction to FI-term is zero and  the vacuum is stable.

For the case of asymmetric model, if the tree level FI term is
\begin{align}
\xi_{{\rm tree}} (y) &= 
\left(
\frac 1 {g q \pi R} \ln\left(\frac {q_\pi}{q_0} \right) + \xi^{\prime \prime} \partial^2_y 
\right)\delta(y)
+\left(
\frac {-1} {g q \pi R} \ln\left(\frac {q_\pi}{q_0} \right) +\xi^{\prime \prime} \partial^2_y \right)
\delta(y-\pi R),
\end{align}
where $\xi''$ is an arbitrary constant satisfying $\xi'' q >0$,
then $\Sigma$ develops its VEV to cancel $\xi_{\rm tree}$.
Substituting $\sigma=gq\xi_0$ in (\ref{eq:xi_asym2}) to the above coefficients of $\delta(y)$,
we can check
$\xi_{{\rm 1-loop}}$ is zero.
The FI-term is stable under quantum corrections.
The vacuum is stable.

In these two cases, the bulk zero mode is localized at the two fixed points and its charge cancels that of the brane fields exactly.
We conclude that such a configuration is the true vacuum of this theory.
It is similar to electrostatic shielding of conductors.

One may suspect that our result depends on regularizations of the delta function.
The cancellation of the 1-loop corrections to the FI-term is a consequence of the localization of the bulk zero mode.
The zero mode function is formally solved as
\begin{align}
\phi_\pm^0 = N \exp\left[\pm \frac {gq \xi_0} 2 y \pm gq\xi''( \delta(y) +\delta(y-\pi R) )\right],
\label{eq:formal}
\end{align}
where $N$ is a normalization factor.
The form of (\ref{eq:formal}) includes infinite series of the products of the delta functions.
It is ill defined, and we need some regularizations.
In general, the result is regularization-dependent, but since $\delta(y)$ diverges at $y=0$, any regularization
of the delta function $\delta_\rho(0)$ diverges as $\rho \rightarrow 0$.
Thus, its divergence at the fixed points is independent of the regularization.

For the massive modes, the Schr\"odinger equation (\ref{eq:wave_equation}) has the potential term $\braket{\Sigma}^2$.
It includes $\delta(y)^2$ and we need some regularizations of the delta function, too.
$\int \delta^2(y)$ (or frequently used regularizations such as Gaussian, sine functions as well as (\ref{eq:reg})) can 
not be normalized and wave functions can not tunnel this potential.
The bulk massive modes are apart from the fixed points.
The quantum corrections to the localized FI-term is dominated by the zero mode.
Therefore, our result would be independent of regularization and it is quite general.

\subsection{$ q + q_0 +q_\pi \neq 0$}
\label{ssec:3.4}

Finally we would like to comment about models with non-vanishing $U(1)$ charge.
If $ q + q_0 +q_\pi \neq 0$, the 4d FI-term is introduced,
\begin{align}
\int_0^{\pi R} \xi(y) dy =\xi_{4{\rm d}} \neq 0. 
\end{align}
If there is a scalar field whose $U(1)$ charge is opposite to the sign of $\xi_{4{\rm d}}$,
it would be possible to cancel the 4d FI-term by its VEV.
The effective FI-term can be set to zero. We may have D-flat moduli space.
Although gauge symmetry is spontaneously broken down,
it would not spoil our previous discussion.
The spontaneous breaking of the gauge symmetry induces masses
for hypermultiplet and brane multiplet,
but it just shifts eigenvalues of their mode expansion,  

If such a scalar does not exist, there is no possibility to absorb the 4d FI-term.
SUSY is broken down and there is no symmetries protecting the coincidence of the coefficients of $D^3$ and $\partial_y \Sigma$.
Our discussion is no longer valid.
We need more considerations for it.

\section{Conclusion}

We have studied quantum corrections to FI-terms in 5 dimensional SUSY Abelian gauge theory compactified on the $S^1/Z_2$ orbifold.
If there is a matter field which is charged under the $U(1)$ symmetry,
the localized FI-term is induced by quantum corrections.
However, the whole story is not so simple.
The localized FI-term changes D-flat condition and $\braket{\Sigma}$.
It affects mode expansion of the bulk fields.
There is no reason to believe that 
the new mode expansion does not induce another FI-term.
We explicitly computed the KK-reduction of the bulk scalar fields with various 
FI-terms and recompute the quantum corrections.
We have shown that the 1-loop correction  vanishes for the symmetric case, but it is not true for the asymmetric case.
It is because the zero mode profile and brane charges cancel each others for the former case, but it does not happen for the latter case.
Therefore, the asymmetric vacuum receives further corrections and is unstable.
If we put a specific FI-term at the tree level Lagrangian, we can realize a stable configuration even for asymmetric case.
In such stable configurations, zero mode profile shields the brane charges completely
and their corrections are canceled.
It would be the true vacuum configurations.
We would like to emphasize that this cancellation is not obtained in the limit of $\Lambda$ to $\infty$, but obtained by finite $\Lambda$.
It is a marked contrast to the result of \cite{GrootNibbelink:2002wv}.

It is important to comment about the supergravity interpretations/extensions of our results. 
As mentioned in the introduction, FI-terms can appear at the orbifold fixed points also in supergravity, even if it is not associated with the $U(1)_R$ symmetry. 
Such localized FI-terms can be considered as the boundary completion of a bulk term~\cite{Barbieri:2002ic}. 
The explicit formulations in 5-dimensional supergravity~\cite{Abe:2004yk,Correia:2004pz} show that the profile~$\xi(y)$ is restricted to the one satisfying $\int_o^{\pi R}dy\,\xi(y)=\xi_{4{\rm d}}=0$, that is, the effective FI-term vanishes in 4-dimensions as it should be. 
Our analyses have been performed for this restricted~$\xi(y)$, which would correspond to the rigid limit of such an FI-term in supergravity. 
Hence, the results of this paper will be applied/extended to supergravity models.\footnote{
On the other hand, in the case with $\xi_{4{\rm d}} \ne 0$ which is mentioned in the subsection~\ref{ssec:3.4}, the gauged $U(1)_R$ symmetry will be required in the supergravity side. At the tree-level, the corresponding FI-term was also formulated and studied in 5-dimensional supergravity~\cite{Abe:2004nx}.}

Phenomenological implication of our result may be drastic and interesting.
Profiles of wave functions in the compact space plays an important role for calculating 4d low energy effective theory.
For example, if quarks and leptons are obtained as chiral zero modes of bulk fields, their zero modes must be localized at the fixed points and its distribution is proportional to the brane charges.
Their overlap integral would have very specific forms.
It may behave like the Froggatt-Nielsen model \cite{Froggatt:1978nt}.
It is important to extend our analysis to $T^2/Z_N$ orbifolds.
From the theoretical perspective, compactifications using toroidal orbifolds with magnetic fluxes is more interesting \cite{Abe:2008fi} 
because it has a clear stringy origin \cite{Bachas:1995ik} and it can realize three families of quarks and leptons as well as its Yukawa hierarchy \cite{Abe:2008sx,Abe:2012fj,Abe:2014vza}.
We shall investigate it elsewhere.

\section*{Acknowledgement }

H. A. is supported by Institute for Advanced Theoretical and Experimental Physics, Waseda University.
T.~K. was  supported in part by MEXT KAKENHI Grant Number JP19H04605.

\appendix

\section{Wave Function of $\phi_-$ for Symmetric Case}
\label{app:odd}

The parity odd solution is given by
\begin{align}
\psi_{-} (y) =
\begin{cases}
	A_- (e^{-\alpha_-y} -e^{-\alpha_+ y}) ~~~&(0 \leq y \leq \rho),\\
	C_- e^{i\sqrt{\lambda}y}+D_- e^{-i\sqrt{\lambda}y} &(\rho < y<\pi R -\rho ), \\
	E_-( e^{\alpha_+ (y-\pi R)}-  e^{\alpha_- (y-\pi R)} )&(\pi R -\rho \leq y \leq \pi R).
\end{cases}
\end{align}
The coefficients are determined by the boundary conditions and normalization.
The boundary condition is written as
\begin{align}
&A_- (e^{-\alpha_- \rho} - e^{-\alpha_+ \rho}) 
= C_- e^{i\sqrt{\lambda} \rho} + D_- e^{-i\sqrt{\lambda} \rho},
\nonumber
\\
&- A_- (\alpha_- e^{-\alpha_{-} \rho} - \alpha_+  e^{-\alpha_{+} \rho}) 
= i\sqrt{\lambda} C_- e^{i\sqrt{\lambda} \rho} -i\sqrt{\lambda}  D_- e^{-i\sqrt{\lambda} \rho},
\nonumber
\\
& C_- e^{i\sqrt{\lambda} (\pi R-\rho)} + D_- e^{-i\sqrt{\lambda} (\pi R-\rho)}
= E_- ( e^{-\alpha_+ \rho} - e^{-\alpha_- \rho}),
\nonumber
\\
& i\sqrt{\lambda} C_- e^{i\sqrt{\lambda} (\pi R-\rho)} -i\sqrt{\lambda}  D_- e^{-i\sqrt{\lambda} (\pi R-\rho)}
= E_-( \alpha_+ e^{-\alpha_+ \rho} - \alpha_- e^{-\alpha_- \rho}).
\end{align}
The eigenvalue $\lambda$ is given by
\begin{align}
\left(\frac{Y+i\sqrt{\lambda}}{Y-i\sqrt{\lambda}}\right)^2
e^{-2i\sqrt{\lambda}(\pi R -2\rho)} =1.
\end{align}
The eigenvalue approaches to $\frac{n^2 }{R^2}$ as $\rho\rightarrow 0$.
The solution with 
$n=0$ is incorrect.
Substituting $\lambda$ to zero, $C_- + D_- =0$ and all the other coefficients is zero.
We have no normalizable solution for $n=0$.
For $n>0$, the boundary conditions are solved:
\begin{align}
A_- &=  \frac{-2i\sqrt{\lambda}}
{(e^{-\alpha_+ \rho} - e^{-\alpha_- \rho})(Y+i\sqrt{\lambda})}
e^{-i\sqrt{\lambda}\rho} D_-,
\nonumber
\\
C_- &= - \frac {Y 
- i \sqrt{\lambda}
}{Y
+i\sqrt{\lambda}
} e^{-2 i \sqrt{\lambda}\rho} D_- ,
\nonumber
\\
E_- &= \frac{-2i\sqrt{\lambda}}
{(e^{-\alpha_+ \rho} - e^{-\alpha_- \rho})(Y-i\sqrt{\lambda})}
e^{-i\sqrt{\lambda}(\pi R - \rho)} D_-,
\end{align}
where 
\begin{align}
Y=\frac {\alpha_+ e^{-\alpha_+ \rho} - \alpha_- e^{-\alpha_- \rho}}
{e^{-\alpha_+ \rho} - e^{-\alpha_- \rho}}.
\end{align}
As $\rho \rightarrow 0$, $Y$ is approaching to zero.
$A_-$ is approaching to $-2D_-$, and $E_-$ is approaching to $-2 (-1)^n D_-$.
$C_-$ is approaching to $ D_-$.
Massive wave functions of $\phi^-$ are approaching to $\cos(ny/R)$ on the bulk of the orbifold.
Around the fixed points,
(\ref{eq:eta}) leads exponential suppression for parity odd modes.
Thus the massive modes of $\phi_-$ can not penetrate to the fixed points, either.
As the result, we obtain wave function profiles of $\phi_-$:
\begin{align}
\phi_-^n (y) =
\begin{cases}
\sqrt{\frac{2}{\pi R}} \cos (\frac {n} R y) &(0<y<\pi R)\\
0 &(y=0,\pi R)
\end{cases}
\end{align}

\section{$\xi_I'' = 0$}
\label{app:xi''=0}

Here, we consider wave function profiles in the case of FI-term without derivative terms:
\begin{align}
\xi = \xi_0 \delta(y) -\xi_0 \delta(y-\pi R).
\end{align}
In this case the wave equation is solved in \cite{GrootNibbelink:2002qp}.
\begin{align}
\phi_{+}^0 &= \sqrt{\frac{2 \sigma}{(e^{2\sigma R}-1)}}\exp\left[\frac 1 2 g q \xi_0(\int_0^y {\rm sgn}(y')dy'- y)\right],
\nonumber
\\
\phi_{+}^n&= \sqrt{\frac 2{\pi R}} \exp\left[\frac 1 2 g q \xi_0(\int_0^y {\rm sgn}(y')dy'- y)\right] \sin(\frac n R y -\theta_n),
\nonumber
\\
\phi_{-}^n&= \sqrt{\frac 2{\pi R}} \exp\left[-\frac 1 2 g q \xi_0(\int_0^y {\rm sgn}(y')dy'- y)\right] \sin(\frac n R y ).
\end{align}
where $n$ is any positive integer.
These wave functions are defined on whole of the compact space since the equation of motion does not have derivative terms of the delta function.
$\phi_-$ has no zero mode.
The eigenvalue $\lambda_n$ and the phase $\theta_n$ are given by
\begin{align}
\lambda_n= \frac{(gq\xi_0)^2}{4}+\frac{n^2}{R^2},~e^{i\theta_n}= \frac{-\frac 1 2 gq \xi_0+i\frac n R}{|-\frac 1 2 gq \xi_0+i\frac n R|}.
\end{align}
The sum of the massive wave functions is computed as
\begin{align}
(|\phi_{+}^n|^2 -|\phi_{-}^n|^2) 
= \frac {2}  {\pi R  \lambda_n}
\left\{
\frac{n^2}{R^2}
\cos(\frac{2n}{R}y)
 + \sigma \frac n R \sin(\frac{2n}{R}y)
\right\} .
\label{eq:sum_n>0}
\end{align}
Thus,  it contributes to the derivative part of FI-term as
\begin{align}
\sum_{n>0} \frac 1 4 \ln\Lambda^2 \lambda_n (|\phi_{+}^n|^2 - |\phi_{-}^n|^2) 
&= -\ln\Lambda^2 \frac {1}{8 \pi R} (\partial_y^2 +2 \sigma \partial_y )\left\{
 \delta(y) + \delta (y-\pi R)
\right\}.
\end{align}
We obtain nonzero correction to the derivative terms of the FI-term.
In addition, the $\Lambda^2$ term is calculated.
\begin{align}
\Lambda^2 \sum_{n>0}(|\phi_{+}^n|^2 -|\phi_{-}^n|^2)  
&= 
\Lambda^2 \sum_{n>0}
\frac {2}  {\pi R  \lambda_n}
\left\{
\frac{n^2}{R^2}
\cos(\frac{2n}{R}y)
 + \sigma \frac n R \sin(\frac{2n}{R}y).
\right\}
\nonumber
\\
&= \frac {\Lambda^2} \pi \partial_y \sum_{n>0} \frac 1 {(\sigma R)^2 +n^2} \left(
n \sin(\frac{2n}{R} y) -\sigma R \cos(\frac{2n}{R} y)
\right).
\label{eq:Lambda}
\end{align}
Using a Fourier transformation formula, the following equation can be computed as
\begin{align}
\frac {a e^{a y}}{e^a -1} =1 +\sum_{n=1}^{\infty} \frac {2a}{a^2 +(2n \pi)^2} \left(
a \cos(\frac {2n}{R} y) -2n\pi \sin(\frac {2n}{R}y) 
\right),
\end{align}
for $0<y<\pi R$.
Then (\ref{eq:Lambda}) is calculated as
\begin{align}
\Lambda^2 \sum_{n>0}(|\phi_{+}^n|^2 -|\phi_{-}^n|^2)
&= \frac {\Lambda^2} {2\sigma R} \partial_y (1-\frac{2\sigma \pi R e^{2\sigma y}}{e^{2\sigma \pi R}-1})
\nonumber
\\
&= -\frac {2\sigma e^{2 \sigma y}}{e^{2\sigma \pi R} -1} \Lambda^2.
\end{align}
In the bulk region, the zero mode profile is calculated as
\begin{align}
\phi_{+}^0 &= \sqrt{\frac{2 \sigma}{(e^{2\sigma R}-1)}}e^{\sigma y}.
\end{align}
Thus, the zero mode and the massive modes cancels each others.
At the fixed points (\ref{eq:Lambda}) diverges again.
We obtain the localized FI-term.

\end{document}